# Axiomatic Quantification of Co-authors' Relative Contributions

Ge Wang[1] and Jiansheng Yang[2]
[1]VT-WFU School of Biomedical Engineering, Virginia Tech, Blacksburg, VA 24061, USA
[2]LMAM, School of Mathematics, Peking University, Beijing, 100871, P.R. China

**Over the past decades, the competition for academic resources has gradually intensified, and worsened with the current financial crisis. To optimize the resource allocation, individualized assessment of research results is being actively studied but the current indices, such as the number of papers, the number of citations, the *h*-factor and its variants have limitations, especially their inability of determining co-authors' credit shares fairly. Here we establish an axiomatic system and quantify co-authors' relative contributions. Our methodology avoids subjective assignment of co-authors' credits using the inflated, fractional or harmonic methods, and provides a quantitative tool for scientific management such as funding and tenure decisions.**

Citation analysis | impact | co-authors' relative contributions | axiomatic approach

**\body**

## Introduction

Because the number of publications and the number of co-authors have been rapidly increasing annually [1], there is a critical and immediate need for individualized assessment of scientific productivity and impact [2-8]. A recent topic in bibliometrics is the use and extension of the *h*-index (defined as the maximum *h* if *h* of a researcher's papers have at least *h* citations each) [4-5] for measurement of his or her academic calibre. While the idea is insightful and widely used [9-13], the *h*-index is quite rough by definition [14] and subject to various biases [15-24]. A major obstacle to significant improvement of the *h*-index and other popular indices of this type has been the lack of a sound mechanism for assessment of co-authors' individual contributions [23, 25].

Current perception of a researcher's qualification relies, to a great degree, on either inflated or fractional counting methods [26-27]. While the former method gives the full credit to any co-author (for example, it is only stated in a biography how many papers are published), the latter method distributes an equally divided recognition to each co-author (as in some bibliometric analyses). Neither of these methods is ideal, because the order or rank of co-authors and the corresponding authorship are almost exclusively used to indicate co-authors' relative contributions. Generally speaking, the further down the list of co-authors for a publication, the less credit he or she receives. Often times, the first author and the corresponding author are considered the most prominent. Now and then, a number of co-authors claim equal contributions.

To quantify co-authors' relative contributions, the harmonic counting method was proposed [27] in order to avoid the equal-share bias of the fractional counting method (a less sophisticated variant was also suggested [8]). While the harmonic counting method does permit equal rankings for subsets of co-authors, without loss of generality let us assume that the order of co-authors is consistent with their credit ranking, and that there are totally $n$ co-authors on a publication whose shares are presented as a vector $\vec{x}=(x_1,\ x_2,\ \cdots,\ x_n)$ ( $1\le i\le n$ ). Then, the *k*-th author's harmonic credit $x_k$ is defined as

$$x_k=\alpha\frac{1}{k}\text{, where } \alpha=\frac{1}{\sum_{j=1}^{n}\frac{1}{j}},\ 1\le k\le n. \qquad (1)$$

Despite its superiority to the fractional method, the harmonic method has not been practically used, due to its subjective nature. Evidently, there is no rationale behind the proportionality that the *k*-th author contributes 1/*k* as much as the first author's contribution. Realistically, there are many possible ratios between the *k*-th and the first authors' credits, which may be equal or may be rather small such as in the cases of data sharing or technical assistance.

Rigorous quantification of co-authors' credits is a long overdue task. The Higher Education Funding Council for England (HEFCE) recently proposed the peer-review system "*Research Excellence Framework (REF)*" that will extensively utilize citation analyses (http://www.nature.com/news/2009/090923/full/news.2009.933.html). Nevertheless, HEFCE has admitted that bibliometrics is not "*sufficiently robust*" for assessment of research quality. Thus, it could be prone to misconducts if those bibliometric measures are administratively used for funding and tenure decisions. For example, a popular Chinese web forum "*New Threads*" (http://www.xys.org/new.html) discussed some cases in which the number of publications, the number of co-authors, and even the *h*-indices were purposely manipulated and effectively inflated. In the USA, the National Institutes of Health recently adopted the enhanced review criteria (http://grants.nih.gov/grants/guide/notice-files/NOT-OD-09-024.html), with mandatory quantification of an investigator's qualification on a 9-point scale (revised from the initially planned 7-point scale). However, the scoring has been largely subjective, still accommodating a substantial level of peer-review noise.

## Results and Discussion

Here we propose to use the axiomatic approach for quantification of co-authors' relative contributions. Assume that a publication has a total of $n$ co-authors who can be divided into $m$ groups ( $n\ge m$ ) and that $c_i$ co-authors in the *i*-th group have the same credit

$x_i \in \vec{x} = (x_1,\ x_2,\ \cdots,\ x_m)$ ( $1 \le i \le m$ ). We postulate the following axioms:

**Axiom 1 (Ranking Preference):**

$x_1 \ge x_2 \ge \cdots \ge x_m > 0$;

**Axiom 2 (Credit Normalization):**

$c_1 x_1 + c_2 x_2 + \cdots c_m x_m = 1$;

**Axiom 3 (Maximum Entropy):** *$\vec{x}$ is uniformly distributed in the domain defined by Axioms 1 and 2.*

The first axiom reflects the ranking process of co-authors' relative contributions, which happens during the production of a publication. In most cases, such a ranking determines the order of co-authors. More efforts beyond this ranking to specify co-authors' credits may well be too complicated, highly controversial, and thus impractical and counter-productive. While a co-authors' contribution statement has been encouraged by some journals, often times it cannot be directly translated into co-authors' credit shares and disappears in the bibliometric measurement. Hence, we suggest that a ranking code be added to each publication as shown in Figure 1, which will be the basis for further analysis. This straightforward ranking code is immediately superior to the inflated and fractional counting methods, since it clearly represents relative importance of co-authors' essential intellectual and technical contributions from their peers' perspectives, and suppresses artifacts in terms of insignificant co-authors, un-qualified corresponding authors, and confusing weights associated with some particular co-authors' positions on a publication [28].

The second axiom ensures that the quantification of co-authors' contributions is in a relative sense. The absolute value of a publication should be estimated independently, which can be the impact factor of a journal initially and the number of citations or its variants subsequently.

The last axiom recognizes the impossibility of specifying exact relative contributions of co-authors on each and every publication, thereby asserting that all the cases permitted by Axioms 1 and 2 are equally likely, since there is no ground for assuming otherwise in the fields of science and technology as a whole. A co-author may have done his or her ultra best for academic excellence or may have only met a minimum requirement, and any scenario in between is quite possible. As in many areas involving information theoretic inference, the maximum entropy principle [29] in this bibliometric context requires that the distribution of the credit vector be uniform across the permissible domain. Nevertheless, in a specific area we could have more information or a stronger assumption. In such a case, our generic axiomatic system can be adapted to make use of available knowledge without any theoretical difficulty.

Therefore, the fairest estimation of co-authors' credit shares can be formulated as the expectation of all possible credit vectors. In other words, the *k*-th set of co-authors' individual credit should be the elemental mean, which is referred to as the *a*-index for its axiomatic foundation and we have proved to be

$$E(x_k) = \frac{1}{m} \sum_{j=k}^{m} \frac{1}{\left( \sum_{i=1}^{j} c_i \right)},\ 1 \le k \le m. \qquad (2)$$

It can be verified that $\frac{1}{m} \sum_{k-1}^{m} c_k E(x_k) = 1$. In the special case of unequal-contribution co-authors (no equal contributions are claimed by any sub-group of these co-authors), Eq. (2) becomes

$$E(x_k) = \frac{1}{n} \sum_{j=k}^{n} \frac{1}{j},\ 1 \le k \le n, \qquad (3)$$

as computed in Table 1 for *n* up to 10.

Our axiomatic characterization is significantly different from the existing credit counting methods. As shown in Figure 2, the fractional measures are too rough compared to the harmonic and axiomatic measures. As far as the harmonic and axiomatic measures are concerned, the axiomatic method promotes the first author's share and dilutes the last author's weight more than the harmonic method does. It is interesting to note that this "Mathew effect" is not only generally desirable but also axiomatically justified.

## Conclusion

We anticipate our axiomatic system to become a basis for development of academic assessment or peer-review systems [23]. It is hoped that our methodology will be adopted by academic institutions and funding agencies, and help improve identification of productive and influential investigators and institutions. Furthermore, our work might be relevant in psychological, social and other contexts in which ranking is fundamentally involved, such as subjective choices and fuzzy reasoning.

## Materials and Methods

Mathematically, our axiomatic quantification problem is to compute not only the elemental mean $E(x_k)$ as a co-author's credit share but also the corresponding standard deviation $\sigma(x_k)$ ( $1 \le k \le m$ ) for statistical testing. The formulas for the co-authors' contributions and the corresponding standard deviations can be derived using either an algebraic or geometric approach. The derivation processes are quite technical, and given in the SI text using the algebraic approach, leading to Eqs. (2) and (3) presented above.

**Acknowledgements**: The authors thank Dr. Zhenjie Lin for discussion and help with implementing a Monte-Carlo algorithm for evaluation of co-authors' credits in this axiomatic framework and Dr. Michael Vannier for advice on the implication and refinement of this work.

## References

1. Greene, M., The demise of the lone author. Nature, 2007. 450(7173): p. 1165.
2. Foulkes, W. and N. Neylon, Redefining authorship. Relative contribution should be given after each author's name. BMJ, 1996. 312(7043): p. 1423.
3. Campbell, P., Policy on papers' contributors. Nature, 1999. 399(6735): p. 393.
4. Hirsch, J.E., An index to quantify an individual's scientific research output. Proc Natl Acad Sci U S A, 2005. 102(46): p. 16569-72.
5. Hirsch, J.E., Does the H index have predictive power? Proc Natl Acad Sci U S A, 2007. 104(49): p. 19193-8.
6. Anonymous, Who is accountable? Nature, 2007. 450(7166): p. 1.
7. Ball, P., A longer paper gathers more citations. Nature, 2008. 455(7211): p. 274-5.
8. Zhang, C.T., A proposal for calculating weighted citations based on author rank. EMBO Rep, 2009. 10(5): p. 416-7.
9. Ball, P., Index aims for fair ranking of scientists. Nature, 2005. 436(7053): p. 900.
10. Ball, P., Achievement index climbs the ranks. Nature, 2007. 448(7155): p. 737.
11. Kinney, A.L., National scientific facilities and their science impact on nonbiomedical research. Proc Natl Acad Sci U S A, 2007. 104(46): p. 17943-7.
12. Pilc, A., The use of citation indicators to identify and support high-quality research in Poland. Arch Immunol Ther Exp (Warsz), 2008. 56(6): p. 381-4.
13. Sebire, N.J., H-index and impact factors: assessing the clinical impact of researchers and specialist journals. Ultrasound Obstet Gynecol, 2008. 32(7): p. 843-5.
14. Dodson, M.V., Citation analysis: Maintenance of h-index and use of e-index. Biochem Biophys Res Commun, 2009. 387(4): p. 625-6.
15. Lehmann, S., A.D. Jackson, and B.E. Lautrup, Measures for measures. Nature, 2006. 444(7122): p. 1003-4.
16. Jeang, K.T., Impact factor, H index, peer comparisons, and Retrovirology: is it time to individualize citation metrics? Retrovirology, 2007. 4: p. 42.
17. Kelly, C.D. and M.D. Jennions, H-index: age and sex make it unreliable. Nature, 2007. 449(7161): p. 403.
18. Wendl, M.C., H-index: however ranked, citations need context. Nature, 2007. 449(7161): p. 403.
19. Engqvist, L. and J.G. Frommen, The h-index and self-citations. Trends Ecol Evol, 2008. 23(5): p. 250-2.
20. Mishra, D.C., Citations: rankings weigh against developing nations. Nature, 2008. 451(7176): p. 244.
21. Radicchi, F., S. Fortunato, and C. Castellano, Universality of citation distributions: toward an objective measure of scientific impact. Proc Natl Acad Sci U S A, 2008. 105(45): p. 17268-72.
22. Todd, P.A. and R.J. Ladle, Citations: poor practices by authors reduce their value. Nature, 2008. 451(7176): p. 244.
23. Bornmann, L. and H.D. Daniel, The state of h index research. Is the h index the ideal way to measure research performance? EMBO Rep, 2009. 10(1): p. 2-6.
24. Baldock, C., R. Ma, and C.G. Orton, Point/counterpoint. The h index is the best measure of a scientist's research productivity. Med Phys, 2009. 36(4): p. 1043-5.
25. Williamson, J.R., My h-index turns 40: my midlife crisis of impact. ACS Chem Biol, 2009. 4(5): p. 311-3.
26. Tscharntke, T., et al., Author sequence and credit for contributions in multiauthored publications. PLoS Biol, 2007. 5(1): p. e18.
27. Hagen, N.T., Harmonic allocation of authorship credit: source-level correction of bibliometric bias assures accurate publication and citation analysis. PLoS One, 2008. 3(12): p. e4021.
28. Laurance, W.F., Second thoughts on who goes where in author lists. Nature, 2006. 442(7098): p. 26.
29. Jaynes, E.T., On the rationale of maximum-entropy methods. Proceedings of the IEEE, 1982. 70(9): p. 939-952.

**Figure Legends**

Fig. 1. Ranking code after the key words to remove any ambiguity in ranking co-authors.

Fig. 2. Comparative visualization of co-authors' relative contributions according to (A) the fractional, (B) harmonic and (C) axiomatic measures respectively, for the number of co-authors up to n=5.

**Table Legend**

Table 1. Axiomatic indices (*a*-indices) for up to 10 unequal-contribution co-authors. Note that the sum of the rounding errors has been added to the first author's share for *n*=4, 5, 8, 9 and 10 respectively.

775 Med. Phys. 27 (4), April 2000 0094-2405/2000/27(4)/775/3/$17.00 © 2000 Am. Assoc. Phys. Med. 775

**Localization of cochlear implant electrodes in radiographs**

Siying Yang
*Department of Electrical & Computer Engineering, The University of Iowa, Iowa City, Iowa 52242*

Ge Wang[a)]
*Department of Radiology, The University of Iowa, Iowa City, Iowa 52242*

Margaret W. Skinner
*Department of Otolaryngology-Head & Neck Surgery, Washington University School of Medicine, Saint Louis, Missouri 63110*

Jay T. Rubinstein
*Department of Otolaryngology, The University of Iowa, Iowa City, Iowa 52242*

Michael W. Vannier
*Department of Radiology, The University of Iowa, Iowa City, Iowa 52242*

Key words: cochlear implantation, x-ray radiography, image processing, thresholding, segmentation

**Ranking code: 1, 2, 3, 3, 2**

Fig. 1. Ranking code after the key words to remove any ambiguity in ranking co-authors.

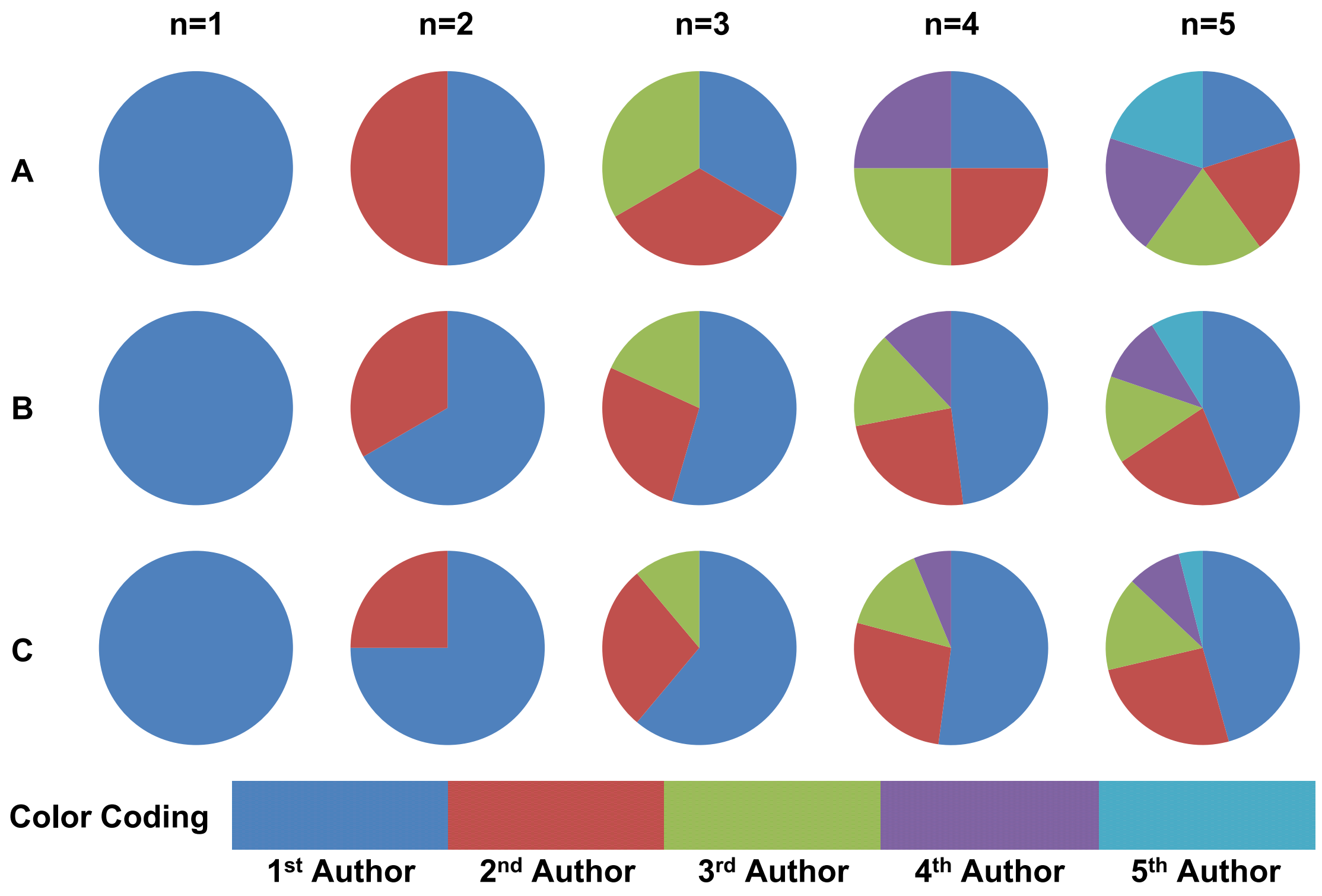

Fig. 2. Comparative visualization of co-authors' relative contributions according to (A) the fractional, (B) harmonic and (C) axiomatic measures respectively, for the number of co-authors up to n=5.

Table 1. Axiomatic indices (*a*-indices) for up to 10 unequal-contribution co-authors. Note that the sum of the rounding errors has been added to the first author's share for $n$=4, 5, 8, 9 and 10 respectively.

| | Co-authors' relative contributions | | | | | | | | | |
|---|---|---|---|---|---|---|---|---|---|---|
| 1 | 1.0000 | | | | | | | | | |
| 2 | 0.7500 | 0.2500 | | | | | | | | |
| 3 | 0.6111 | 0.2778 | 0.1111 | | | | | | | |
| 4 | 0.5209 | 0.2708 | 0.1458 | 0.0625 | | | | | | |
| 5 | 0.4566 | 0.2567 | 0.1567 | 0.0900 | 0.0400 | | | | | |
| 6 | 0.4083 | 0.2417 | 0.1583 | 0.1028 | 0.0611 | 0.0278 | | | | |
| 7 | 0.3704 | 0.2276 | 0.1561 | 0.1085 | 0.0728 | 0.0442 | 0.0204 | | | |
| 8 | 0.3398 | 0.2147 | 0.1522 | 0.1106 | 0.0793 | 0.0543 | 0.0335 | 0.0156 | | |
| 9 | 0.3145 | 0.2032 | 0.1477 | 0.1106 | 0.0828 | 0.0606 | 0.0421 | 0.0262 | 0.0123 | |
| 10 | 0.2928 | 0.1929 | 0.1429 | 0.1096 | 0.0846 | 0.0646 | 0.0479 | 0.0336 | 0.0211 | 0.0100 |

## SI Methodology: Derivation of a-indices and associated deviations

Ge Wang[1] and Jiansheng Yang[2]
[1]VT-WFU School of Biomedical Engineering, Virginia Tech, Blacksburg, VA 24061, USA
[2]LMAM, School of Mathematics, Peking University, Beijing, 100871, P.R. China

As described in the main text, there are totally $n$ co-authors on a publication who can be divided into $m$ groups ($n \geq m$), and $c_i$ co-authors in the *i*-th group have the same credit $x_i \in \vec{x} = (x_1,\ x_2,\ \cdots,\ x_m)$ ($1 \leq i \leq m$). Our axiomatic system consists of

**Axiom 1 (Ranking Preference):** $x_1 \geq x_2 \geq \cdots \geq x_m > 0$;

**Axiom 2 (Credit Normalization):** $c_1 x_1 + c_2 x_2 + \cdots c_m x_m = 1$;

**Axiom 3 (Maximum Entropy):** $\vec{x}$ *is uniformly distributed in the domain defined by Axioms 1 and 2.*

Then, our problem is to compute not only the elemental mean $E(x_k)$ as a co-author's credit share but also the corresponding standard deviation $\sigma(x_k)$ ($1 \leq k \leq m$) for statistical testing. For visualization of the key idea, the 3D case is illustrated in Figure S1.

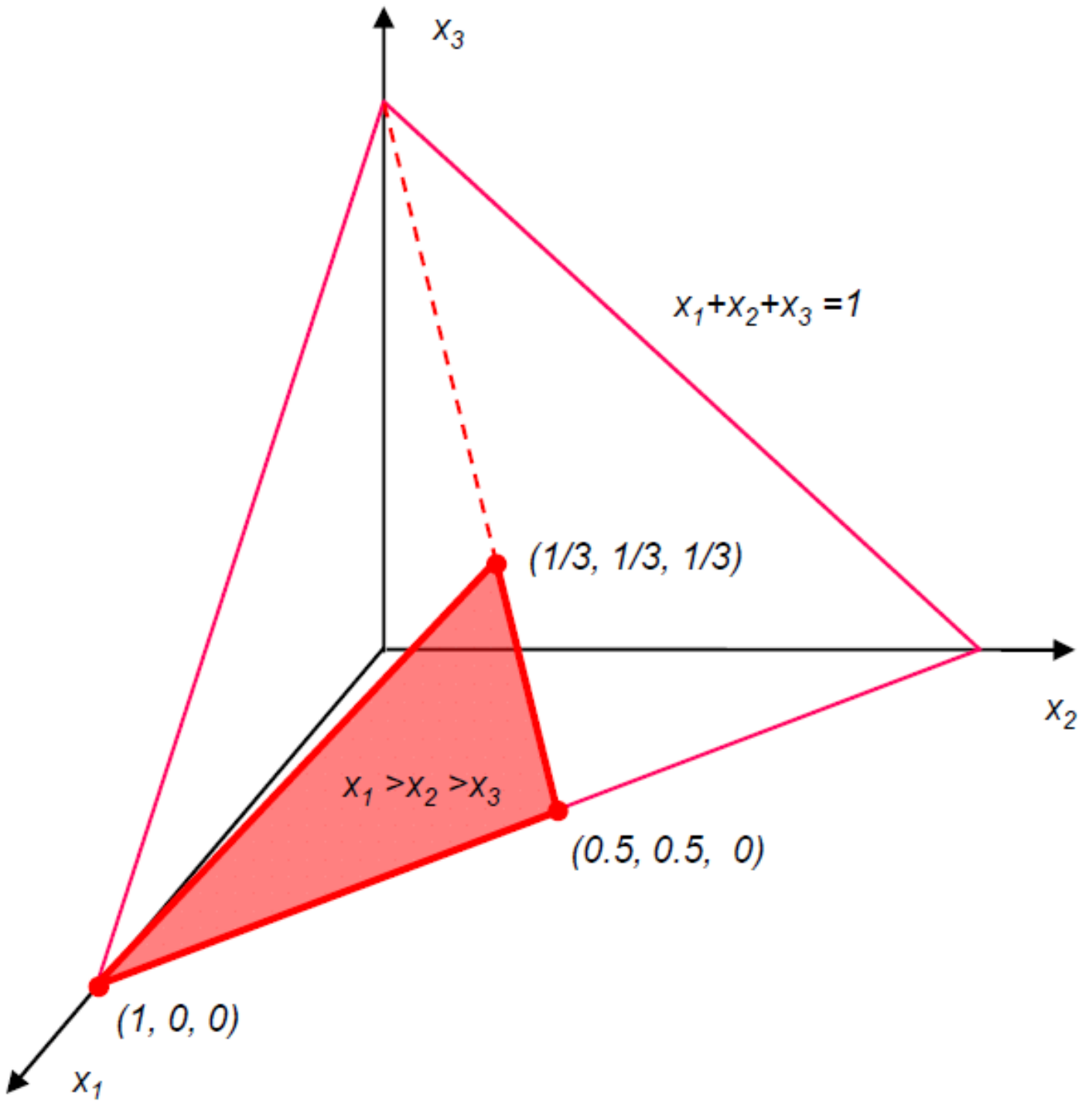

Figure S1. Domain permitted by the axiomatic system in the case of n=3, where the distribution of co-authors' credit shares is postulated to be the mass center of the solid red triangle.

Since $\sigma(x_k) = \sqrt{E(x_k{}^2) - E(x_k)^2}$ , we will need to find $E(x_k{}^2)$ ($1 \leq k \leq m$). For convenience of the induction to be used below, let us denote $E(x_k)$ and $E(x_k{}^2)$ ($1 \leq k \leq m$) by $R_{m,k}$ and $S_{m,k}$ respectively.

The sample space of the above problem is

$$\Omega_k = \{ \bar{x}(m) = (x_2, \cdots x_m) : 0 \leq x_m \leq x_{m-1} \leq \cdots \leq x_2 \leq \frac{1}{c_1}\left(1 - \sum_{i=2}^{m} c_i x_i\right) \} . \qquad (1)$$

Let

$$M_m = \int_{\Omega_m} d\overline{x}(m) \,, \tag{2}$$

$$E_{m,k} = \int_{\Omega_m} x_k d\overline{x}(m) \,, \; 1 \le k \le m \,, \tag{3}$$

$$F_{m,k} = \int_{\Omega_m} {x_k}^2 d\overline{x}(m) \,, \; 1 \le k \le m \,, \tag{4}$$

where

$$x_1 = \frac{1}{c_1}\left(1 - \sum_{i=2}^{m} c_i x_i\right). \tag{5}$$

Clearly, we have

$$R_{m,k} = \frac{E_{m,k}}{M_m} \,, \; 1 \le k \le m \,, \tag{6}$$

$$S_{m,k} = \frac{F_{m,k}}{M_m} \,, \; 1 \le k \le m \,. \tag{7}$$

To determine $R_{m,k}$ and $S_{m,k}$ $(1 \le k \le m)$ in a recursive fashion, we introduce the following functions whose utilities will become evident later:

$$\overline{\Omega}_m(a,b) = \{\overline{x}(m) = (x_2, \cdots x_m) : b \le x_m \le x_{m-1} \le \cdots \le x_2 \le \frac{1}{c_1}\left(a - \sum_{i=2}^{m} c_i x_i\right)\} \,, \tag{8}$$

$$\overline{M}_m(a,b) = \int_{\overline{\Omega}_m(a,b)} d\overline{x}(m) \,, \tag{9}$$

$$\overline{E}_{m,k}(a,b) = \int_{\overline{\Omega}_m(a,b)} x_k d\overline{x}(m) \,, \; 1 \le k \le m \,, \tag{10}$$

$$\overline{F}_{m,k}(a,b) = \int_{\overline{\Omega}_m(a,b)} {x_k}^2 d\overline{x}(m) \,, \; 1 \le k \le m \,. \tag{11}$$

where

$$x_1 = \frac{1}{c_1}\left(a - \sum_{i=2}^{m} c_i x_i\right), \tag{12}$$

and $a$ and $b$ are constants with $a \ge 0$ , $a \ge b\sum_{i=1}^{m} c_i$ . Then, we have the following propositions.

**Proposition 1 (Equivalency)**: $\overline{M}_m(a,b) = \overline{M}_m(a - b\sum_{i=1}^{m} c_i, 0)$ ;

**Proposition 2 (Measurement)**: $\overline{M}_m(a,0) = \dfrac{a^{m-1}}{(m-1)!(c_1+c_2)(c_1+c_2+c_3)\cdots(c_1+c_2+\cdots+c_m)}$ ;

**Proposition 3 (Reduction)**: For $1 \le k \le m$ , we have

$$\bar{E}_{m,k}(a,b)=\left(a-b\sum_{i=1}^{m}c_i\right)^m E_{m,k}+\frac{b\left(a-b\sum_{i=1}^{m}c_i\right)^{m-1}}{(m-1)!(c_1+c_2)(c_1+c_2+c_3)\cdots(c_1+c_2+\cdots+c_m)}\ ;$$

$$\bar{F}_{m,k}(a,b)=\left(a-b\sum_{i=1}^{m}c_i\right)^{m+1} F_{m,k}+2b\left(a-b\sum_{i=1}^{m}c_i\right)^m E_{m,k}+\frac{b^2\left(a-b\sum_{i=1}^{m}c_i\right)^{m-1}}{(m-1)!(c_1+c_2)(c_1+c_2+c_3)\cdots(c_1+c_2+\cdots+c_m)};$$

**Proposition 4 (Substitution)**: $R_{2,1}=\frac{1}{c_1}(1-c_2R_{2,2})$ , $S_{2,1}=\frac{1}{{c_1}^2}(1-2c_2R_{2,2}+{c_2}^2S_{2,2})$ .

**Proof of Proposition 1:** Making the variables transform $y_i=x_i-b$ for $2\le i\le m$ , we have

$$\bar{M}_m(a,b)=\int_{\bar{\Omega}_m(a,b)}d\bar{x}(m)=\int_{\bar{\Omega}_m(a-b\sum_{i=1}^{m}c_i,0)}d\bar{y}(m)=\bar{M}_m(a-b\sum_{i=1}^{m}c_i,0)\,. \tag{13}$$

**Proof of Proposition 2:** Let us proceed by induction with respect to $m$ . For $m=2$ , we have

$$\bar{M}_2(a,0)=\int_{\bar{\Omega}_2(a,0)}d\bar{x}(2)=\int_0^{\frac{a}{c_1+c_2}}dx_2=\frac{a}{c_1+c_2}\,. \tag{14}$$

That is , the proposition holds in this case.

For $m>2$ , by **Proposition 2** and the inductive hypothesis, we have

$$\bar{M}_m(a,0)=\int_{\bar{\Omega}_m(a,0)}d\bar{x}(m)$$

$$=\int_0^{\frac{a}{c_1+c_2+\cdots+c_m}}\left(\int_{\bar{\Omega}_{m-1}(a-c_mx_m,x_m)}d\bar{x}(m-1)\right)dx_m$$

$$=\int_0^{\frac{a}{c_1+c_2+\cdots+c_m}}\left(\int_{\bar{\Omega}_{m-1}(a-x_m\sum_{i=1}^{m}c_i,0)}d\bar{x}(m-1)\right)dx_m$$

$$=\int_0^{\frac{a}{c_1+c_2+\cdots+c_m}}\bar{M}_{m-1}(a-x_m\sum_{i=1}^{m}c_i,0)dx_m$$

$$=\int_0^{\frac{a}{c_1+c_2+\cdots+c_m}}\frac{(a-x_m\sum_{i=1}^{m}c_i)^{m-2}}{(m-2)!(c_1+c_2)(c_1+c_2+c_3)\cdots(c_1+c_2+\cdots+c_{m-1})}dx_m$$

$$=\frac{a^{m-1}}{(m-1)!(c_1+c_2)(c_1+c_2+c_3)\cdots(c_1+c_2+\cdots+c_m)}\,. \tag{15}$$

**Proof of Proposition 3:** For $1\le k\le m$ , we have

$$\bar{E}_{m,k}(a,b)=\int_{\bar{\Omega}_m(a,b)}(x_k-b)d\bar{x}(m)+\int_{\bar{\Omega}_m(a,b)}bd\bar{x}(m)$$

$$=\int_{\bar{\Omega}_m(a,b)}(x_k-b)d\bar{x}(m)+b\bar{M}_m(a,b)\,, \tag{16}$$

$$\bar{F}_{m,k}(a,b)$$

$$=\int_{\bar{\Omega}_m(a,b)}(x_k-b)^2d\bar{x}(m)+\int_{\bar{\Omega}_m(a,b)}2b(x_k-b)d\bar{x}(m)+\int_{\bar{\Omega}_m(a,b)}b^2d\bar{x}(m)$$

$$=\int_{\bar{\Omega}_m(a,b)}(x_k-b)^2d\bar{x}(m)+2b\int_{\bar{\Omega}_m(a,b)}(x_k-b)d\bar{x}(m)+b^2\bar{M}_m(a,b)\,. \tag{17}$$

Making the variables transform $y_i=\dfrac{x_i-b}{a-b\sum_{i=1}^{m}c_i}$ for $1\le i\le m$, we have

$$\int_{\bar{\Omega}_m(a,b)}(x_k-b)d\bar{x}(m)=\left(a-b\sum_{i=1}^{m}c_i\right)^m\int_{\Omega_m}y_kd\bar{y}(m)=\left(a-b\sum_{i=1}^{m}c_i\right)^m E_{m,k}\,, \tag{18}$$

$$\int_{\bar{\Omega}_m(a,b)}(x_k-b)^2d\bar{x}(m)=\left(a-b\sum_{i=1}^{m}c_i\right)^{m+1}\int_{\Omega_m}{y_k}^2d\bar{y}(m)=\left(a-b\sum_{i=1}^{m}c_i\right)^{m+1} F_{m,k}\,, \tag{19}$$

By **Propositions 1 and 2**, we have

$$\bar{M}_m(a,b)=\frac{\left(a-b\sum_{i=1}^{m}c_i\right)^{m-1}}{(m-1)!(c_1+c_2)(c_1+c_2+c_3)\cdots(c_1+c_2+\cdots+c_m)}\,. \tag{20}$$

Inserting Eqs. (18), (19) and (20) into Eqs. (16) and (17) respectively, we obtain

$$\bar{E}_{m,k}(a,b)=\left(a-b\sum_{i=1}^{m}c_i\right)^m E_{m,k}+\frac{b\left(a-b\sum_{i=1}^{m}c_i\right)^{m-1}}{(m-1)!(c_1+c_2)(c_1+c_2+c_3)\cdots(c_1+c_2+\cdots+c_m)}\,, \tag{21}$$

$$\bar{F}_{m,k}(a,b)=\left(a-b\sum_{i=1}^{m}c_i\right)^{m+1} F_{m,k}+2b\left(a-b\sum_{i=1}^{m}c_i\right)^m E_{m,k}+$$

$$+\frac{b^2\left(a-b\sum_{i=1}^{m}c_i\right)^{m-1}}{(m-1)!(c_1+c_2)(c_1+c_2+c_3)\cdots(c_1+c_2+\cdots+c_m)}. \tag{22}$$

**Proof of Proposition 4:** For $m=2$, we have

$$c_1x_1+c_2x_2=1, \tag{23}$$

$$x_1=\frac{1}{c_1}(1-c_2x_2)\,. \tag{24}$$

Hence, we have

$$R_{2,1} = E(x_1) = E(\frac{1}{c_1}(1-c_2x_2)) = \frac{1}{c_1}E(1-c_2x_2)$$

$$= \frac{1}{c_1}(E(1)-c_2E(x_2)) = \frac{1}{c_1}(1-c_2R_{2,2}), \quad (25)$$

and

$$S_{2,1} = E(x_1{}^2) = E(\frac{1}{c_1{}^2}(1-c_2x_2)^2) = \frac{1}{c_1{}^2}E((1-c_2x_2)^2)$$

$$= \frac{1}{c_1{}^2}(E(1)-2c_2E(x_2)+c_2{}^2E(x_2{}^2))$$

$$= \frac{1}{c_1{}^2}(1-2c_2R_{2,2}+c_2{}^2S_{2,2}). \quad (26)$$

**Theorem 1:** $R_{m,k} = \frac{1}{m}\sum_{j=k}^{m}\frac{1}{c_1+\cdots+c_j}$, $1 \le k \le m$.

**Proof:** By **Propositions 1 and 2**, we have

$$E_{m,m} = \int_{\Omega_m} x_m d\bar{x}(m) = \int_0^{\frac{1}{c_1+c_2+\cdots+c_m}} x_m \left(\int_{\bar{\Omega}_{m-1}(1-c_mx_m,x_m)} d\bar{x}(m-1)\right) dx_m$$

$$= \int_0^{\frac{1}{c_1+c_2+\cdots+c_m}} x_m \bar{M}_{m-1}(1-c_mx_m, x_m) dx_m$$

$$= \int_0^{\frac{1}{c_1+c_2+\cdots+c_m}} x_m \bar{M}_{m-1}(1-x_m\sum_{i=1}^{m}c_i, 0) dx_m$$

$$= \int_0^{\frac{1}{c_1+c_2+\cdots+c_m}} \frac{x_m\left(1-x_m\sum_{i=1}^{m}c_i\right)^{m-2}}{(m-2)!(c_1+c_2)(c_1+c_2+c_3)\cdots(c_1+c_2+\cdots+c_{m-1})} dx_m$$

$$= \frac{1}{(c_1+c_2+\cdots+c_m)m!(c_1+c_2)(c_1+c_2+c_3)\cdots(c_1+c_2+\cdots+c_m)}. \quad (27)$$

By **Proposition 3** and Eq. (27), for $1 \le k \le m-1$ we have

$$E_{m,k}(c_1,\cdots,c_m) = \int_{\Omega_m(c_1,\cdots,c_m)} x_k d\bar{x}(m) = \int_0^{\frac{1}{c_1+c_2+\cdots+c_m}} \left(\int_{\bar{\Omega}_{m-1}(1-c_mx_m,x_m,c_1,\cdots,c_{m-1})} x_k d\bar{x}(m-1)\right) dx_m$$

$$= \int_0^{\frac{1}{c_1+c_2+\cdots+c_m}} \bar{E}_{m-1,k}(1-c_mx_m, x_m, c_1,\cdots,c_{m-1}) dx_m$$

$$= \int_0^{\frac{1}{c_1+c_2+\cdots+c_m}} \left( (1-x_m\sum_{i=1}^{m}c_i)^{m-1} E_{m-1,k}(c_1,c_2,\cdots,c_{m-1}) + \frac{x_m(1-x_m\sum_{i=1}^{m}c_i)^{m-2}}{(m-2)!(c_1+c_2)(c_1+c_2+c_3)\cdots(c_1+c_2+\cdots+c_{m-1})} \right) dx_m$$

$$= E_{m,m} + \frac{E_{m-1,k}}{m(c_1+c_2+\cdots+c_m)} \ . \tag{28}$$

By **Proposition 2**, we have

$$M_m = \bar{M}_m(1.0) = \frac{1}{(m-1)!(c_1+c_2)(c_1+c_2+c_3)\cdots(c_1+c_2+\cdots+c_m)} \ . \tag{29}$$

Hence, we have

$$R_{m,m} = \frac{E_{m,m}}{M_m} = \frac{1}{m(c_1+\cdots+c_m)} \ , \tag{30}$$

and for $1 \le k \le m-1$ we have

$$R_{m,k} = \frac{E_{m,k}}{M_m} = \frac{E_{m,m}}{M_m} + \frac{E_{m-1,k}}{m(c_1+c_2+\cdots+c_m)M_m}$$

$$= R_{m,m} + \frac{R_{m-1,k}M_{m-1}}{m(c_1+c_2+\cdots+c_m)M_m}$$

$$= R_{m,m} + \frac{m-1}{m}R_{m-1,k} \ . \tag{31}$$

For $2 \le k \le m-1$, repeatedly using Eq. (31) we have

$$R_{m,k} = R_{m,m} + \frac{m-1}{m}(R_{m-1,m-1} + \frac{m-2}{m-1}R_{m-2,k})$$

$$= R_{m,m} + \frac{m-1}{m}R_{m-1,m-1} + \frac{m-2}{m}R_{m-2,k}$$

$$= R_{m,m} + \frac{m-1}{m}R_{m-1,m-1} + \frac{m-2}{m}R_{m-2,m-2} + \cdots + \frac{k}{m}R_{k,k}$$

$$= \frac{1}{m(c_1+\cdots+c_m)} + \frac{1}{m(c_1+\cdots+c_{m-1})} + \frac{1}{m(c_1+\cdots+c_{m-2})} + \cdots \frac{1}{m(c_1+\cdots+c_k)}$$

$$= \frac{1}{m}\sum_{j=k}^{m}\frac{1}{c_1+\cdots+c_j} \ . \tag{32}$$

For $k=1$, repeatedly using Eq. (31) we have

$$R_{m,1} = R_{m,m} + \frac{m-1}{m}R_{m-1,m-1} + \frac{m-2}{m}R_{m-2,m-2} + \cdots + \frac{3}{m}R_{3,3} + \frac{2}{m}R_{2,1}$$

$$= \frac{1}{m}\sum_{j=3}^{m}\frac{1}{c_1+\cdots+c_j} + \frac{2}{m}R_{2,1} \ . \tag{33}$$

Inserting $R_{2,1}=\frac{1}{c_1}(1-c_2R_{2,2})=\frac{1}{c_1}(1-\frac{c_2}{2(c_1+c_2)})$ into Eq. (33), we have

$$R_{m,1}=\frac{1}{m}\sum_{j=3}^{m}\frac{1}{c_1+\cdots+c_j}+\frac{2}{mc_1}(1-\frac{c_2}{2(c_1+c_2)})=\frac{1}{m}\sum_{j=1}^{m}\frac{1}{c_1+\cdots+c_j}. \quad (34)$$

Combining Eqs. (30), (32) and (34), we have

$$R_{m,k}=\frac{1}{m}\sum_{j=k}^{m}\frac{1}{c_1+\cdots+c_j},\ 1\le k\le m. \quad (35)$$

**Theorem 2:** $S_{m,k}=\frac{2}{m(m+1)}\sum_{k\le i\le j\le m}\frac{1}{(c_1+\cdots+c_j)(c_1+\cdots+c_i)}$ , $1\le k\le m$.

**Proof:** By **Propositions 1 and 2**, we have

$$F_{m,m}=\int_{\Omega_m}x_m{}^2d\overline{x}(m)=\int_0^{\frac{1}{c_1+c_2+\cdots+c_m}}x_m{}^2\left(\int_{\overline{\Omega}_{m-1}(1-c_mx_m,x_m)}d\overline{x}(m-1)\right)dx_m$$

$$=\int_0^{\frac{1}{c_1+c_2+\cdots+c_m}}x_m{}^2\overline{M}_{m-1}(1-c_mx_m,x_m)dx_m$$

$$=\int_0^{\frac{1}{c_1+c_2+\cdots+c_m}}x_m{}^2\overline{M}_{m-1}(1-x_m\sum_{i=1}^{m}c_i,0)dx_m$$

$$=\int_0^{\frac{1}{c_1+c_2+\cdots+c_m}}\frac{x_m{}^2\left(1-x_m\sum_{i=1}^{m}c_i\right)^{m-2}}{(m-2)!(c_1+c_2)(c_1+c_2+c_3)\cdots(c_1+c_2+\cdots+c_{m-1})}dx_m$$

$$=\frac{2}{(c_1+c_2+\cdots+c_m)^2(m+1)!(c_1+c_2)(c_1+c_2+c_3)\cdots(c_1+c_2+\cdots+c_m)}. \quad (36)$$

For $1\le k\le m-1$, utilizing **Proposition 3** and Eq. (36), we have

$$F_{m,k}=\int_{\Omega_m}x_k{}^2d\overline{x}(m)=\int_0^{\frac{1}{c_1+c_2+\cdots+c_m}}\left(\int_{\overline{\Omega}_{m-1}(1-c_mx_m,x_m)}x_k{}^2d\overline{x}(m-1)\right)dx_m$$

$$=\int_0^{\frac{1}{c_1+c_2+\cdots+c_m}}\overline{F}_{m-1,k}(1-c_mx_m,x_m)dx_m$$

$$=\int_0^{\frac{1}{c_1+c_2+\cdots+c_m}}\left(\left(1-x_m\sum_{i=1}^{m}c_i\right)^{m}F_{m-1,k}+2x_m\left(1-x_m\sum_{i=1}^{m}c_i\right)^{m-1}E_{m-1,k}\right)dx_m$$

$$+\int_0^{\frac{1}{c_1+c_2+\cdots+c_m}}\frac{x_m{}^2\left(1-x_m\sum_{i=1}^{m}c_i\right)^{m-2}}{(m-2)!(c_1+c_2)(c_1+c_2+c_3)\cdots(c_1+c_2+\cdots+c_{m-1})}dx_m$$

$$=F_{m,m}+\frac{F_{m-1,k}}{(m+1)(c_1+\cdots+c_m)}+\frac{2E_{m-1,k}}{m(m+1)(c_1+\cdots+c_m)^2}, \quad (37)$$

Therefore, we have

$$S_{m,m} = \frac{F_{m,m}}{M_m} = \frac{2}{m(m+1)(c_1 + \cdots + c_m)^2}\ , \tag{38}$$

and for $1 \le k \le m-1$ we have

$$S_{m,k} = \frac{F_{m,m}}{M_m} + \frac{F_{m-1,k}}{(m+1)(c_1 + \cdots + c_m)M_m} + \frac{2E_{m-1,k}}{m(m+1)(c_1 + \cdots + c_m)^2 M_m}$$

$$= S_{m,m} + \frac{m-1}{m+1} S_{m-1,k} + \frac{2(m-1)}{m(m+1)(c_1 + \cdots + c_m)} R_{m-1,k}\ . \tag{39}$$

For $2 \le k \le m-1$, repeatedly using (39) we have

$$S_{m,k} = S_{m,m} + \frac{2(m-1)}{m(m+1)(c_1 + \cdots + c_m)} R_{m-1,k} + \frac{m-1}{m+1} S_{m-1,k}$$

$$= S_{m,m} + \frac{2(m-1)}{m(m+1)(c_1 + \cdots + c_m)} R_{m-1,k}$$

$$+ \frac{m-1}{m+1}\left( S_{m-1,m-1} + \frac{2(m-2)}{(m-1)m(c_1 + \cdots + c_{m-1})} R_{m-2,k} + \frac{m-2}{m} S_{m-2,k} \right)$$

$$= S_{m,m} + \frac{m-1}{m+1} S_{m-1,m-1} + \frac{2(m-1)}{m(m+1)(c_1 + \cdots + c_m)} R_{m-1,k} + \frac{2(m-2)}{m(m+1)(c_1 + \cdots + c_{m-1})} R_{m-2,k}$$

$$+ \frac{(m-1)(m-2)}{m(m+1)} S_{m-2,k}$$

$$= S_{m,m} + \frac{m-1}{m+1} S_{m-1,m-1} + \frac{(m-1)(m-2)}{(m+1)m} S_{m-2,m-2} + \cdots + \frac{(k+1)k}{(m+1)m} S_{k,k}$$

$$+ \frac{2(m-1)}{m(m+1)(c_1 + \cdots + c_m)} R_{m-1,k} + \frac{2(m-2)}{m(m+1)(c_1 + \cdots + c_{m-1})} R_{m-2,k} + \cdots + \frac{2k}{m(m+1)(c_1 + \cdots + c_{k+1})} R_{k,k}$$

$$= \frac{2}{m(m+1)}\left( \sum_{j=k}^{m} \frac{1}{(c_1 + \cdots + c_j)^2} + \sum_{k \le i < j \le m} \frac{1}{(c_1 + \cdots + c_j)(c_1 + \cdots + c_i)} \right)$$

$$= \frac{2}{m(m+1)} \sum_{k \le i \le j \le m} \frac{1}{(c_1 + \cdots + c_j)(c_1 + \cdots + c_i)}\ . \tag{40}$$

For $k = 1$, repeatedly using Eq. (39) we have

$$S_{m,1} = S_{m,m} + \frac{m-1}{m+1} S_{m-1,m-1} + \frac{(m-1)(m-2)}{(m+1)m} S_{m-2,m-2} + \cdots + \frac{4\times 3}{(m+1)m} S_{3,3} + \frac{3\times 2}{(m+1)m} S_{2,1}$$

$$+ \frac{2(m-1)}{m(m+1)(c_1 + \cdots + c_m)} R_{m-1,1} + \frac{2(m-2)}{m(m+1)(c_1 + \cdots + c_{m-1})} R_{m-2,1} + \cdots + \frac{2\times 2}{m(m+1)(c_1 + c_2 + c_3)} R_{2,1}$$

$$= \frac{2}{m(m+1)}\left( \sum_{j=3}^{m} \frac{1}{(c_1+\cdots+c_j)^2} + \sum_{1\le i<j\le m} \frac{1}{(c_1+\cdots+c_j)(c_1+\cdots+c_i)} + 3S_{2,1} - \frac{1}{c_1(c_1+c_2)} \right). \tag{41}$$

Since

$$S_{2,1} = \frac{1}{c_1^{\,2}}(1-2c_2R_{2,2}+c_2^{\,2}S_{2,2}) = \frac{1}{c_1^{\,2}}(1-\frac{2c_2}{2(c_1+c_2)}+\frac{2c_2^{\,2}}{2\times 3(c_1+c_2)^2})$$

$$= \frac{1}{c_1^{\,2}} - \frac{c_2}{c_1^{\,2}(c_1+c_2)} + \frac{c_2^{\,2}}{3c_1^{\,2}(c_1+c_2)^2} = \frac{c_1+c_2-c_2}{c_1^{\,2}(c_1+c_2)} + \frac{c_2^{\,2}}{3c_1^{\,2}(c_1+c_2)^2}$$

$$= \frac{1}{c_1(c_1+c_2)} + \frac{c_2^{\,2}}{3c_1^{\,2}(c_1+c_2)^2}, \tag{42}$$

we have

$$3S_{2,1} - \frac{1}{c_1(c_1+c_2)} = \frac{c_2^{\,2}}{c_1^{\,2}(c_1+c_2)^2} + \frac{2}{c_1(c_1+c_2)} = \frac{c_2^{\,2}+2c_1(c_1+c_2)}{c_1^{\,2}(c_1+c_2)^2}$$

$$= \frac{(c_1+c_2)^2+c_1^{\,2}}{c_1^{\,2}(c_1+c_2)^2} = \frac{1}{c_1^{\,2}} + \frac{1}{(c_1+c_2)^2}. \tag{43}$$

Inserting Eq. (43) into Eq. (41), we have

$$S_{m,1} = \frac{2}{m(m+1)}\left( \sum_{j=1}^{m} \frac{1}{(c_1+\cdots+c_j)^2} + \sum_{1\le i<j\le m} \frac{1}{(c_1+\cdots+c_j)(c_1+\cdots+c_i)} \right). \tag{44}$$

Combining Eqs. (36), (40) and (44), we obtain

$$S_{m,k} = \frac{2}{m(m+1)} \sum_{k\le i\le j\le m} \frac{1}{(c_1+\cdots+c_j)(c_1+\cdots+c_i)}, \quad \text{for } 1\le k\le m. \tag{45}$$

Finally, we have

**Theorem 3:** For $1\le k\le m$,

$$\sigma(x_k) = \frac{1}{m}\sqrt{\frac{m-1}{m+1}\sum_{j=k}^{m}\frac{1}{(c_1+\cdots+c_j)^2} - \frac{2}{m+1}\sum_{k\le i<j\le m}\frac{1}{(c_1+\cdots+c_j)(c_1+\cdots+c_i)}}.$$

**Proof:** For $1\le k\le m$, we have

$$\sigma(x_k) = \sqrt{E(x_k^{\,2})-E(x_k)^2} = \sqrt{S_{m,k}-R_{m,k}^{\,2}}$$

$$= \frac{1}{m}\sqrt{\frac{m-1}{m+1}\sum_{j=k}^{m}\frac{1}{(c_1+\cdots+c_j)^2} - \frac{2}{m+1}\sum_{k\le i<j\le m}\frac{1}{(c_1+\cdots+c_j)(c_1+\cdots+c_i)}}. \tag{46}$$

**Remark:** In the case of $m=n$, $c_1=c_2=\cdots=c_m=1$, we have

$$E(x_k) = \frac{1}{n}\sum_{j=k}^{n}\frac{1}{j},$$

$$\sigma(x_k) = \frac{1}{n}\sqrt{\frac{n-1}{n+1}\sum_{j=k}^{n}\frac{1}{j^2} - \frac{2}{n+1}\sum_{k\le i<j\le n}\frac{1}{i\cdot j}},\ 1\le k\le n. \qquad (47)$$